\documentclass[%
 reprint,
superscriptaddress,
 amsmath,amssymb,
 aps,
 prl,
]{revtex4-2}

\usepackage{amsmath}
\allowdisplaybreaks
\usepackage{graphicx}
\usepackage{amssymb, amsfonts, amsmath,framed}
\usepackage{bm}
\usepackage{nicefrac}
\usepackage[colorlinks=true]{hyperref}
\usepackage[dvipsnames]{xcolor}
\hypersetup{
    colorlinks=true,
    allcolors=Violet,
}
\usepackage[shortlabels]{enumitem}
\usepackage[normalem]{ulem}

\newcommand{\dd}{\mathrm{d}}

\newcommand{\bu}{\boldsymbol{u}}

\newcommand{\bQ}{\boldsymbol{Q}}
\newcommand{\bnabla}{\boldsymbol{\nabla}}
\newcommand{\bcdot}{\boldsymbol{\cdot}}

\newcommand{\Rh}{R_{\mathrm{heat}}}
\newcommand{\mcRh}{\mathcal{R}_{\mathrm{heat}}}
\newcommand{\Rped}{R_{\mathrm{ped}}}
\newcommand{\wh}{w_{\mathrm{heat}}}

\newcommand{\mcE}{\mathcal{E}}
\newcommand{\const}{\mathrm{const}}

\newcommand{\p}{\partial}

\newcommand{\mcL}{\mathcal{L}}

\newcommand{\Ma}{\mathrm{Ma}_{\phi}}

\begin{document}

\title{Explosive eruption cycles in a rotating Z-pinch}

\author{David N. Hosking}
\email{dnh26@cam.ac.uk}
\affiliation{Department of Applied Mathematics and Theoretical Physics, University of Cambridge, Centre for Mathematical Sciences, Wilberforce Road, Cambridge, CB3 0WA, UK}
\affiliation{Gonville \& Caius College, Trinity Street, Cambridge, CB2 1TA, UK}
\author{Luca Swinnerton}
\affiliation{Department of Applied Mathematics and Theoretical Physics, University of Cambridge, Centre for Mathematical Sciences, Wilberforce Road, Cambridge, CB3 0WA, UK}
\author{Rahul Kesavan}
\affiliation{Department of Applied Mathematics and Theoretical Physics, University of Cambridge, Centre for Mathematical Sciences, Wilberforce Road, Cambridge, CB3 0WA, UK}

\date{\today}

\begin{abstract}
A transonic shear flow directed along magnetic field lines can linearly stabilize a steep pressure gradient in a confined magnetohydrodynamic (MHD) plasma. In Z-pinch geometry, we show that, like the edge pedestal in tokamak devices, this transport barrier---which we call the ``MHD pedestal''---is \emph{metastable}, i.e., unstable to finite-amplitude displacements of flux tubes. We simulate the slow formation of an MHD pedestal in a heated and sheared Z-pinch, which collapses on reaching a critical height, expelling an order-unity fraction of the confined thermal energy. The MHD pedestal then rebuilds and the process repeats, in a manner analogous to the ELM cycle seen in fusion experiments. We show that the available energy of the metastable equilibrium, and the most energetically favorable amount of ejected plasma, can be calculated from first principles via combinatorial optimization of flux-tube interchanges.
\end{abstract}

\maketitle

\mbox{\textit{Introduction.\quad}} Since experiments on the ASDEX device in the early 1980s, it has been known empirically that, at sufficient plasma-heating power, tokamak fusion experiments enter a reduced-turbulence state known as high-confinement mode (H-mode)~\cite{Wagner82_ASDEX, Wagner84_ASDEX}. H-mode is characterized by steep pressure and density gradients at the plasma edge, called the \emph{pedestal} because—owing to the stiffness of core heat transport—raising the plasma temperature near the edge enables it to be higher in the core~\cite{ITER99_c2, Garbet04_stiffness, Ryter01_stiffness}. This makes operation in H-mode essential to plans for power-plant-scale fusion reactors~\cite{ITER07_c3, Hughes20_SPARC, Lennholm24_STEP, Tholerus24_STEP}.

Given the importance of H-mode, a major challenge for magnetic-confinement fusion is the quasi-periodic relaxation of the pedestal by eruptions of hot plasma filaments. These \emph{edge-localized modes} (ELMs) impose large transient heat loads on plasma-facing components---the thermal energy liberated in a type-I ELM can be~${\sim10\%}$ of that in the device~\cite{Hughes20_SPARC}---and are therefore a central concern for reactor-scale operation~\cite{ConnorKirkWilson08_ELMs}. Considerable progress has been made in predicting the pedestal state at the type-I ELM boundary: the EPED model, which estimates pedestal height and width from the intersection of a peeling-ballooning stability limit with a kinetic-ballooning constraint on pedestal width, has reproduced pedestal parameters at ELM onset to~${\sim20\%}$ accuracy across broad conditions in the Alcator C-Mod and \mbox{DIII-D} tokamaks, and forms a widely used basis for ITER pedestal projections~\cite{Snyder11_EPED, Snyder12_EPED, Walk12, Groebner13_EPED}. However, EPED is a static pedestal model and does not describe the nonlinear crash dynamics. Despite advances in phenomenological modeling~\cite{Zohm96_ELMs, Helander99, Gimblett06_ELMs, Chapman01_sandpile, Chapman01b_sandpile} and simulation~\cite{Cathey20_Jorek}, key aspects of these dynamics remain unclear, including why plasma at the peeling-ballooning stability boundary erupts abruptly, how these eruptions saturate, and whether the released energy can be predicted from first principles.

Sudden eruptions from the pedestal may occur because the peeling–ballooning stability boundary marks the loss of a metastable pedestal state~\cite{CowleyArtun97, CowleyCowley15}. Ref.~\cite{GuazzottoBetti11} demonstrated numerically, using an equilibrium geometry from the DIII-D tokamak, that a transonic shear flow in an MHD plasma can generate an edge transport barrier that they termed the ``MHD pedestal''. Here, we show that a sheared field-aligned plasma flow in Z-pinch geometry produces the same effect, and that the corresponding MHD pedestal is metastable. We simulate a Z-pinch plasma heated slowly to the stability boundary, at which point the flux-tube equilibria are lost in a saddle-node bifurcation, triggering rapid flux-tube expulsion and collapse of the MHD pedestal. The pedestal then rebuilds, and the process repeats cyclically.

\mbox{\textit{Interchange stability in a Z-pinch.\quad}}In this paper, we consider the dynamics of a Z-pinch with ${\beta \equiv 2p/B^2\ll 1}$, where $p$ is the thermal pressure and $B$ the azimuthal magnetic field, with a transonic field-aligned flow. In this case, the large energetic cost of bending field lines requires that the dynamics be axisymmetric~\footnote{The Z-pinch has two canonical ideal-MHD instabilities: the axisymmetric ``sausage'' and the non-axisymmetric ``kink''. The instability criterion for the sausage instability is $-\dd \ln p/\dd \ln r > 2\gamma/(1+\gamma \beta/2)$, where $\gamma$ is the adiabatic index, while that of the kink is $-\dd \ln p/\dd \ln r > m^2/\beta$, where $m\geq 1$ is the azimuthal mode number. The former is most restrictive provided $3\gamma \beta<2$, i.e., $\beta < 2/5$ for $\gamma = 5/3$.}. In cylindrical coordinates $(r,\phi, z)$ with unit vectors $(\boldsymbol{\hat{r}}, \boldsymbol{\hat{\phi}}, \boldsymbol{\hat{z}})$, the poloidal velocity ${\bu_{\mathrm{pol}}\equiv \bu - u_{\phi}\boldsymbol{\hat{\phi}}}$ satisfies
\begin{equation}
    \rho\frac{\dd \bu_{\mathrm{pol}}}{\dd t} = -\bnabla P + f \boldsymbol{\hat{r}},\label{2D_EOM}
\end{equation}where $\dd/\dd t = \p/\p t + \bu_{\mathrm{pol}}\bcdot \bnabla$, $P = p+B^2/2$ is the total pressure, $\rho$ the density, and $f$ the net body force on a circular field line owing to magnetic tension and inertia:
\begin{equation}
    f = -\frac{B^2}{r}+\frac{\rho u_{\phi}^2}{r}.\label{f}
\end{equation}Eq.~\eqref{f} is of the form $f=f(P, r, \bQ)$, where $\bQ$ is a vector whose components are the functionally independent Lagrangian invariants of the plasma: $\bQ = (s, \chi, \ell)$, where $s=p^{1/\gamma}/\rho$ is a specific entropy proxy, $\chi = B/\rho r$ is the specific magnetic flux, and $\ell = u_{\phi} r$ is the specific angular momentum. Written in these variables, with axial current $I\equiv B r=\const$ and $\gamma$ the adiabatic index,  Eq.~\eqref{f} becomes, to leading order in $\beta \ll 1$,
\begin{equation}
    f = -\frac{2P}{r} + \frac{I}{r^5}\left[\frac{2 I^{\gamma-1}}{r^{2\gamma-4}}\left(\frac{s}{\chi}\right)^{\gamma}+\frac{\ell^2}{\chi}\right].\label{flowbeta}
\end{equation}

By analogy with Archimedes' principle, a flux tube moved in local total-pressure balance from $r_1$ to $r_2$ experiences a net radial acceleration 
\begin{multline}
    a_{1\to2} = \frac{f(P_2, r_2, \bQ_1)-f(P_2, r_2, \bQ_2)}{\rho(P_2,r_2,\bQ_1)},
    \label{interchange_force}\\
     = \frac{\chi_1}{r_2^3}\left\{\frac{2I^{\gamma -1 }}{r_2^{2\gamma-4}}\left[\left(\frac{s_1}{\chi_1}\right)^{\gamma}-\left(\frac{s_2}{\chi_2}\right)^{\gamma}\right]+\frac{\ell_1^2}{\chi_1}-\frac{\ell_2^2}{\chi_2}\right\}
\end{multline}where $X_{i}=X_{\mathrm{eq}}(r_i)$, with $X_{\mathrm{eq}}(r)$ the equilibrium profile of $X$. At linear order in ${\delta r = r_2-r_1}$, the condition for $a_{1\to 2}$ to be restoring, i.e., directed oppositely to $\delta r$, is
\begin{equation}
    \mathcal{L}\equiv -\lim_{\delta r\to 0}\frac{a_{1\to2}}{\delta r}= \frac{c_s^2}{r} \left(2\frac{\dd}{\dd r}\ln \frac{s}{\chi} + \Ma^2\frac{\dd }{\dd r}\ln\frac{\ell^2}{\chi}\right) > 0.\label{linear_stability_low_beta}
\end{equation}
where $c_{s}\equiv\sqrt{\gamma p/\rho}$ is the sound speed and $\Ma\equiv u_{\phi}/c_{s}$ is the Mach number of the azimuthal flow.

\mbox{\textit{The MHD pedestal.}\quad} 
Eq.~\eqref{linear_stability_low_beta} shows that a sharp increase in the $\ell^2/\chi$-profile of a transonic flow can linearly stabilize a similarly sharp radial decrease in $s/\chi$. We refer to such a feature as an ``MHD pedestal'', or, for brevity, ``pedestal'', in what follows, in analogy with Ref.~\cite{GuazzottoBetti11}. The height of a sharp pedestal at $r=R_{\mathrm{ped}}$ separating hot, stationary plasma at $r<R_{\mathrm{ped}}$ from cold plasma flowing azimuthally at $r>R_{\mathrm{ped}}$ is limited by stability: According to Eq.~\eqref{interchange_force}, the acceleration of a flux tube displaced from $r_{-}=R_{\mathrm{ped}}-\delta r$ to $r_{+}=R_{\mathrm{ped}}+\delta r$, with $\delta r\to 0$, is restoring if 
\begin{equation}
    \hspace{-1mm}\frac{2I^{\gamma-1}}{\Rped^{2\gamma-4}}\frac{s^\gamma}{\chi^\gamma}\bigg|_{r_{-}}=\frac{\ell^2}{\chi}\bigg|_{r_{+}}\hspace{-1mm}\implies  p(r_{-})< \frac{1}{2}\rho u_{\phi}^2\bigg|_{r_+}.\label{narrowcondition}
\end{equation}Thus, the plasma pressure at the MHD pedestal is limited to half the ram pressure of the confining flow.

 \begin{figure}
    \centering
    \includegraphics[width=1.0\columnwidth]{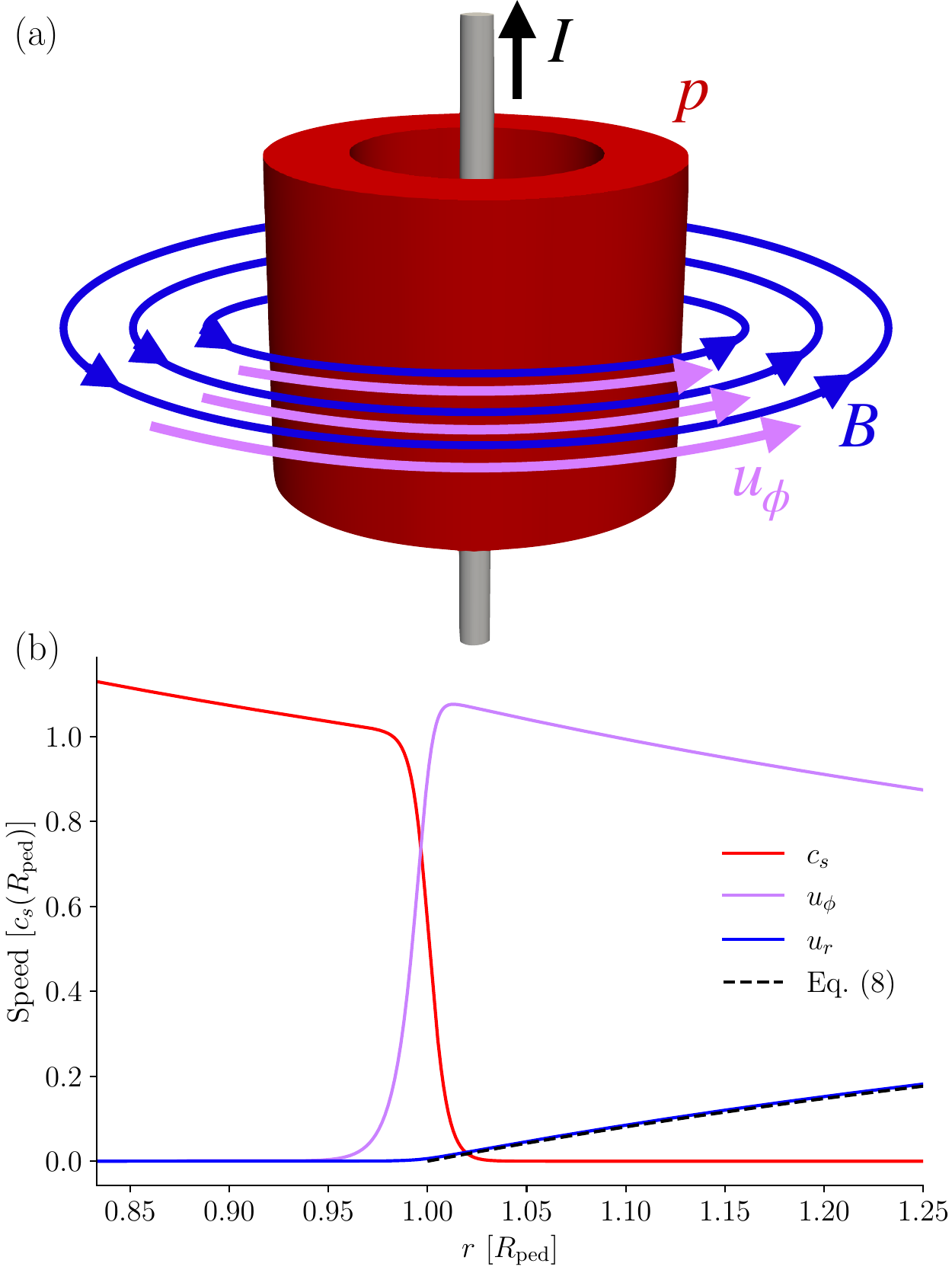}
    \caption{Panel (a): Visualization of a rotating, hard-core Z-pinch. Fast-flowing cold plasma at large radii stabilizes a sharp gradient of thermal pressure, confining hot plasma to the red annulus. Panel (b): Numerical example of an equilibrium at marginal linear stability [$\mcL=0$ in Eq.~\eqref{linear_stability_low_beta}] with a $\tanh$ profile for $\ell(r)$ and $\chi(r)=\const$ (red and pink lines). The blue line shows the velocity of a flux tube erupting from $r<R_{\mathrm{ped}}$ and the black dashed line shows Eq.~\eqref{singlefluxtubemotion}.}
    \label{fig:trajectory_illustration}
\end{figure}

\mbox{\textit{Metastability of the MHD pedestal.}\quad} Although a positive gradient of $\ell^2/\chi$ can stabilize a negative gradient of $s/\chi$ to linear displacements, it cannot stabilize sufficiently large nonlinear displacements (provided $\gamma<2$). This is because, as $r_2\to\infty$, the $s/\chi$ terms in Eq.~\eqref{interchange_force} necessarily dominate the $\ell^2/\chi$ terms. A linearly stable MHD pedestal in a Z-pinch is therefore only \emph{metastable}.

\mbox{\textit{Eruption trajectory at marginal linear stability.}\quad} Both the early- and late-time trajectories of a flux tube erupting from a marginally stable ($\mcL=0$) pedestal are tractable analytically [see Fig.~\ref{fig:trajectory_illustration}(b) for the full trajectory in an example case]. When the flux-tube displacement $\delta r$ is much smaller than the pedestal width (the WKBJ limit), we may Taylor expand Eq.~\eqref{interchange_force} to obtain
\begin{equation}
    \frac{\dd u_r}{\dd t}= a_{1\to 2} = -\mathcal{L} \, \delta r + \mathcal{N}\, \delta r^2\underset{\mathcal{L}=0}{\implies} \delta r = \frac{6}{\mathcal{N}}\frac{1}{(t-t_0)^2}, \label{durdt}
\end{equation}where $t_0$ is an integration constant. We show in the End Matter that ${\mathcal{N}= -2(2-\gamma )I^{\gamma-1}r^{-2\gamma}\dd(s^{\gamma}/\chi^{\gamma})/\dd r}$ if ${\mathcal{L}=0}$ everywhere. Thus, the outwards motion of the flux tube is \emph{explosive} when $s/\chi$ decreases with radius. When the displacement of the flux tube is greater than the pedestal width, we may use Eq.~\eqref{narrowcondition} to yield $\chi_2 r_2^3 a_{1\to2}  = \chi_1\ell(r_{+})^2 [(r_2/\Rped)^{4-2\gamma}-1
]$. Taking $\chi_2$ to be constant, and with $\Gamma \equiv {(2-\gamma)}/{(\gamma -1)}$,
\begin{equation}
    \frac{\rho(r_{-}) u_r(r_2)^2}{\rho(r_{+}) u_{\phi}(r_{+})^2} = \Gamma+\left(\frac{R_{\mathrm{ped}}}{r_2}\right)^{2}- \frac{1}{\gamma - 1}\left(\frac{R_{\mathrm{ped}}}{r_2}\right)^{2\gamma-2}.\label{singlefluxtubemotion}
\end{equation}The velocity $u_r$ grows exponentially (${u_r\propto r_2 - \Rped}$) for ${r_2/R_{\mathrm{ped}}-1\ll 1}$ and approaches a constant as $r_2\to\infty$.

\mbox{\textit{Model forcing.}\quad}A minimal heating and torquing scheme that leads to an MHD pedestal in a Z-pinch is as follows. Starting from a $\beta\ll 1$, $\Ma\sim 1$ reference state for which $\bQ=\bQ_0=\const$ (the subscript zero denotes the reference state in what follows), we evolve the thermal energy and azimuthal velocity according to
\begin{equation}
    \frac{\dd }{\dd t}\frac{p}{\gamma -1 }= -\frac{\gamma}{\gamma-1}p\bnabla \bcdot \bu_{\rm pol} + \mcRh\wh(r)+\mathcal{S}_p,\label{modelthermal}
\end{equation}and
\begin{equation}
    \frac{\dd u_{\phi}}{\dd t} = - \frac{u_{\phi}}{\tau_{\mathrm{drag,\phi}}}w_{\mathrm{in}}(r) - \frac{u_{\phi}-u_{\phi,0}(r)}{\tau_{\mathrm{out}}}w_{\mathrm{out}}(r)
    + \mathcal{S}_{u_\phi},\label{utheta_damping}
\end{equation}respectively, where we have dropped diffusive terms for brevity. Eq.~\eqref{modelthermal} describes heating at rate $\mcRh$ localized to radius $\Rh$ by the window function ${0<\wh(r)<1}$, which has width $\Delta r_{\rm heat}$. Eq.~\eqref{utheta_damping} forces a transonic shear with $\Ma\ll 1$ for ${r<\Rped}$ and $\Ma\simeq 1$ for $r > \Rped$, as in Ref.~\cite{GuazzottoBetti11} [$w_{\mathrm{in}}(r)$ and $w_{\mathrm{out}}(r)$ are step functions]. The terms $\mathcal{S}_p$ and $\mathcal{S}_{u_\phi}$ represent relaxation of $\bQ$ to~$\bQ_0$ on a timescale $\tau_{\mathrm{wall}}$ for $r\gtrsim 2.5\Rh$ (see the Supplementary Information for details).

 \begin{figure}
    \centering
    \includegraphics[width=\columnwidth]{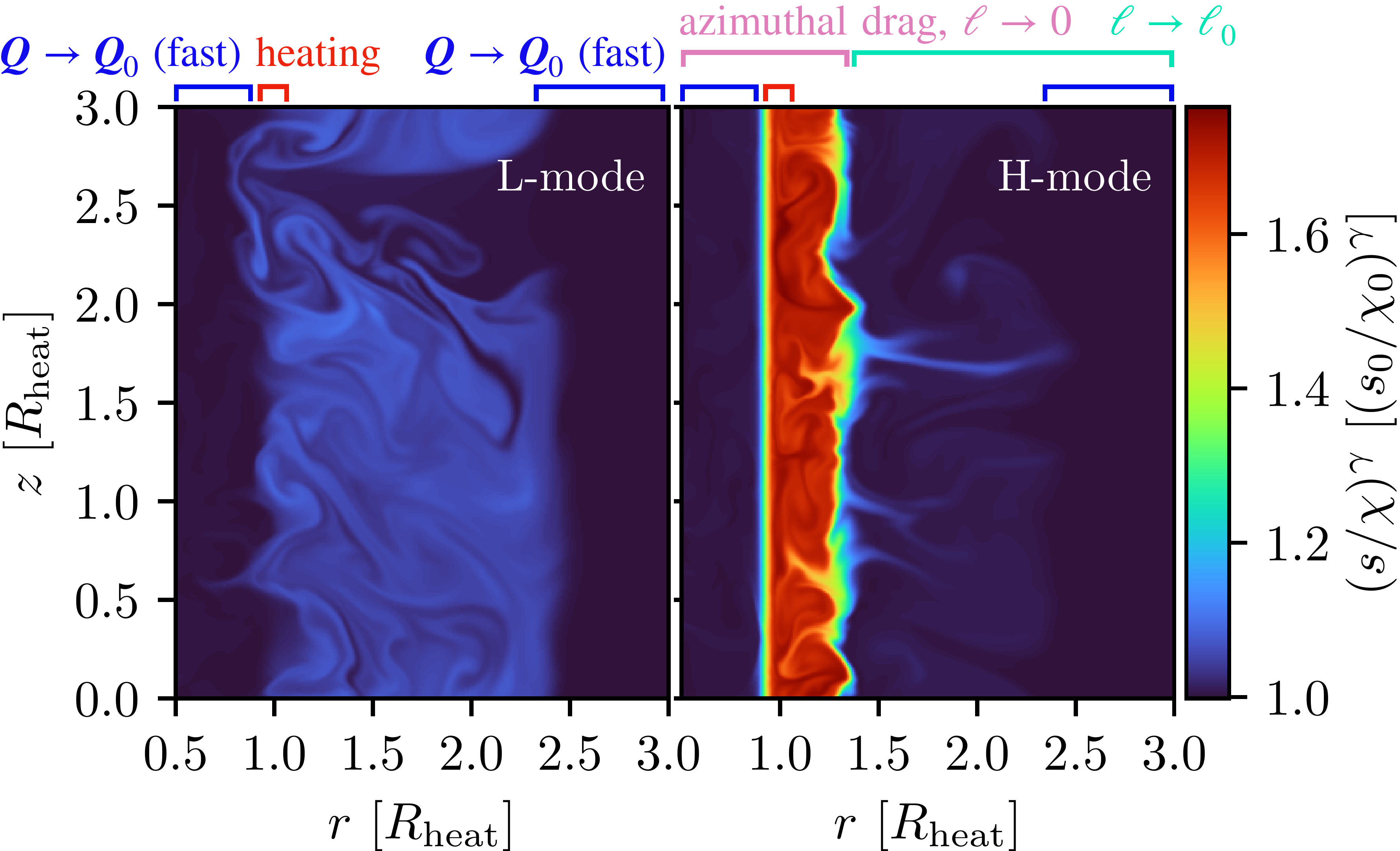}
    \caption{Left panel: Numerical simulation without toroidal flow damping [i.e., $\tau_{\mathrm{drag,\phi}}\to \infty$ in Eq.~\eqref{utheta_damping}]. The plasma is driven to an interchange-marginal state with $s/\chi\sim \const$. Right panel: Numerical simulation with $\tau_{\mathrm{drag,\phi}}\sim \tau_{\mathrm{heat}}$. The plasma is interchange-marginal both within and outside of a hot core, but with a sharp gradient of $s/\chi$ between these regions (the ``MHD pedestal''). Annotations above each panel indicate the spatial localizations of the sources and sinks in Eqs.~\eqref{modelthermal} and~\eqref{utheta_damping}. A movie showing the transition from L-mode to H-mode is available at~\url{https://youtu.be/Ayh1-HcQd54}.}
    \label{fig:LH_comparison}
\end{figure}

\begin{figure*}
    \centering
    \includegraphics[width=\textwidth]{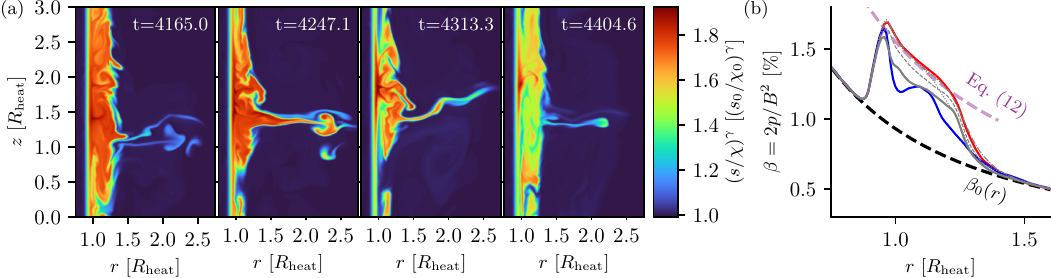}
    \caption{Eruption event in the H-mode simulation shown in the right panel of Fig.~\ref{fig:LH_comparison}, at a later time. The first image in panel~(a) shows the state of the plasma before the eruption; its $\beta=2p/B^2$ profile (averaged over $z$) is shown in red in panel (b). The next three images show, from left to right: a mid-eruption state; a post-eruption state [blue line in panel (b)]; and the post-eruption state after interchange-driven mixing in the core has homogenized the $s/\chi$ distribution there [thick gray line in panel (b)]. The times given are in units of $\Rh/c_{s,0}(\Rh)$. The dashed, dash-dotted and dotted lines show the subsequent rebuild of the pedestal to marginal linear stability, before the next eruption in the cycle. A movie version of this figure is available at \url{https://youtu.be/8U26HgsziPM}.}
    \label{fig:collapse}
\end{figure*}

\mbox{\textit{Turbulence regimes.}\quad} Eqs.~\eqref{modelthermal} and~\eqref{utheta_damping} support two distinct turbulence regimes, with the realized regime determined by the relative magnitudes of the heating and torquing timescales, as explained below. Fig.~\ref{fig:LH_comparison} visualizes the two regimes in numerical simulations conducted with the Athena++ MHD code~\cite{Stone20}. These simulations have $\gamma = 5/3$, ${\Rped = 1.25\Rh}$, ${\Delta r_{\mathrm{heat}}\simeq 0.1 \Rh}$ and ${p_0(\Rh)/\mcRh\simeq 200 \Rh/c_{s,0}(\Rh)}$. The sink timescale is ${\tau_{\mathrm{wall}}\simeq 0.5 \Rh/c_{s,0}(\Rh)}$, which is around $10$ Alfv\'{e}nic box-crossing times, so flux tubes at the sink maintain pressure balance with their surroundings. A conservative drag term $-\rho \bu_{\mathrm{pol}}/\tau_{\mathrm{drag,pol}}$, with ${\tau_{\rm drag,pol}\sim 50\Rh/c_{s,0}(\Rh)}$, on the right-hand side of Eq.~\eqref{2D_EOM} provides an energy sink for the 2D inverse cascade that would otherwise transfer energy to box-scale convective cells~\footnote{Empirically, we find ${\tau_{\rm drag,pol}\sim 50\Rh/c_{s,0}(\Rh)}$ to be small enough to allow convective motions to dissipate between eruption events, while large enough not to suppress the eruptions.}. We present full details of the simulations in the Supplementary Information.

\mbox{\textit{(i) ``L-mode''.}\quad} The left panel of Fig.~\ref{fig:LH_comparison} shows a case with ${\tau_{{\rm drag},\phi} = \tau_{{\rm out}} = \infty}$, i.e., no azimuthal torques. The heating at radius $\Rh$ in Eq.~\eqref{modelthermal} drives an interchange-turbulence-mediated heat flux outward to a sink at larger radius (left panel of Fig.~\ref{fig:LH_comparison}). The interchange turbulence mixes the system to $s/\chi\sim \const$ (${s/\chi\simeq s_0/\chi_0}$ as the sink is aggressive compared with the heating: $p_0(\Rh)/\mcRh \gg \tau_{\mathrm{wall}}$).

\mbox{\textit{(ii) ``H-mode''.}\quad} 
The right panel of Fig.~\ref{fig:LH_comparison} shows a simulation with ${\tau_{\mathrm{drag,\phi}}\simeq \tau_{\mathrm{out}}\simeq 800 \Rh/c_{s,0}(\Rh)}$. The drag at ${r<\Rped}$ produces a stabilizing gradient of angular momentum [see Eq.~\eqref{linear_stability_low_beta}] at $\Rped$, which disrupts the ``L-mode''-like turbulent state and allows an annulus of hotter plasma to develop, with an MHD pedestal at $r=\Rped$. In general, the condition to enter this ``H-mode'' state is that $f(P,r,\bQ)$ decreases (becomes more negative) in a Lagrangian sense as the net result of heating and the azimuthal drag at $r<\Rped$. The characteristic net heating rate of the volume $V_{\rm core}$ defined by $\Rh<r<\Rped$ is
\begin{equation}
    \frac{1}{\tau_{\mathrm{heat}}}\equiv \frac{\dd}{\dd t}\ln\int_{V_{\mathrm{core}}}\dd V\frac{p}{\gamma - 1}\sim \frac{\Delta r_{\mathrm{heat}}}{\Rped-\Rh}\frac{\mcRh}{p_0}.
\end{equation}
Estimating $\p_t f \propto c_{s,0}^2/\tau_{\mathrm{heat}} - u_{\phi,0}^2/\tau_{\mathrm{drag,\phi}}$ from Eq.~\eqref{flowbeta} yields $\tau_{\mathrm{heat}}\gtrsim\tau_{\mathrm{drag,\phi}}$ as the condition to enter H-mode for $u_{\phi,0}^2\sim c_{s,0}^2$ ($\tau_{\mathrm{heat}}\simeq \tau_{\mathrm{drag,\phi}}$ for the right panel of Fig.~\ref{fig:LH_comparison}~\footnote{We find that $\tau_{\mathrm{heat}}\simeq \tau_{\mathrm{drag,\phi}}$ is necessary so that the counterflow that replaces erupted material is not suppressed by a highly stable core. Cases we have examined with $\tau_{\mathrm{heat}}\gg \tau_{\mathrm{drag,\phi}}$ do not exhibit pedestal collapse.}). Using $\chi\simeq \chi_0$ [$\chi$ is not forced by Eqs.~\eqref{modelthermal} and~\eqref{utheta_damping}], the greatest well-mixed value of $s$ that can be reached by the hot annulus while remaining linearly stable is [cf. Eq.~\eqref{narrowcondition}] ${s_H^{\gamma} = s_0^{\gamma} (1+\gamma u_{\phi,0}^2/2 c_{s,0}^2)}|_{\Rped}$, whence the pressure profile at marginal stability is
\begin{equation}
    p_H =  \left[1+\frac{\gamma u_{\phi, 0}(\Rped)^2}{2 c_{s,0}(\Rped)^2}\right] p_0(\Rh) \left(\frac{r}{\Rh}\right)^{-2\gamma}.\label{p_H}
\end{equation}

\mbox{\textit{Eruption and transport-barrier collapse.}\quad} On reaching the stability boundary, Eq.~\eqref{p_H}, the H-mode state erupts abruptly. Fig.~\ref{fig:collapse} visualizes one particular collapse event, which is the ninth such event in the simulation. \color{black}The first of the four panels in Fig.~\ref{fig:collapse}(a) shows the state of the plasma prior to the burst, which broadly resembles the right panel of Fig.~\ref{fig:LH_comparison}. The corresponding pressure profile, normalized by the magnetic pressure (i.e., plotted as the plasma $\beta$) is shown in red in Fig.~\ref{fig:collapse}(b). This is, upon averaging over $z$, coincident with the prediction of Eq.~\eqref{p_H} between $\Rh$ and $R_{\mathrm{ped}}$. The second panel in Fig.~\ref{fig:collapse}(a) visualizes the column mid-eruption, a short time later (around two sound-crossing times of the simulation domain). The panel shows a plume of hot material moving toward the sink. The plume forms a mushroom-shaped head, as is characteristic of the nonlinear phase of Rayleigh--Taylor-like instabilities. The plume visualized is one of many (around $10$) that erupt consecutively from the plasma column in this period. The third panel in Fig.~\ref{fig:collapse}(a) shows the post-eruption state (a weak plume remains at $r>1.5\Rh$), corresponding to the most suppressed pressure profile (blue line in Fig.~\ref{fig:collapse}). Cold plasma from outside the column flows inward to replace the erupted hot plasma. The final stage of the process is the mixing of hot, un-erupted plasma remaining in the core with the cold inflow, facilitated by interchange turbulence in the core. By the time of the final panel in Fig.~\ref{fig:collapse}(a), the core has roughly homogenized its distribution of $s/\chi$, giving a pressure profile parallel but somewhat below the H-mode one, Eq.~\eqref{p_H}. Subsequently, the core reheats, reaching the critical profile around $t=4600\Rh/c_{s,0}(\Rh)$, and triggering a new eruption around $t=4700\Rh/c_{s,0}(\Rh)$.

\begin{figure*}
    \centering
    \includegraphics[width=\textwidth]{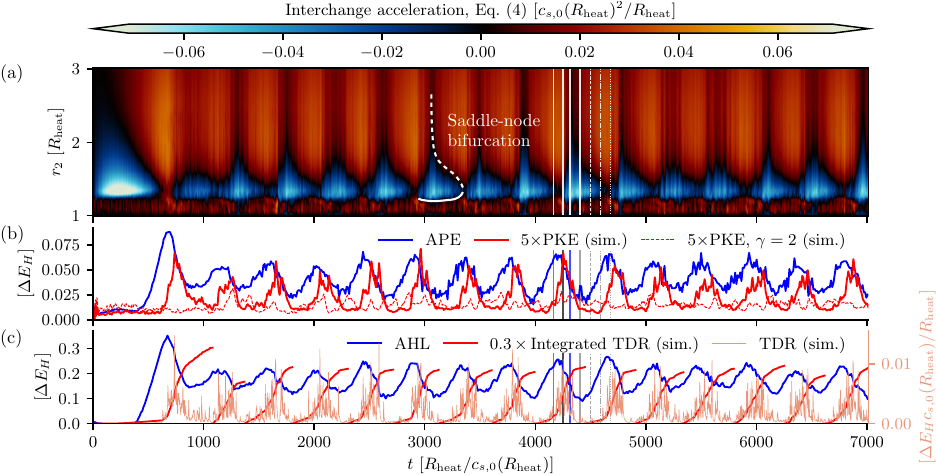}
    \caption{Panel (a): Interchange acceleration [Eq.~\eqref{interchange_force}] as a function of the displaced flux tube's radius $r_2$. Eruptions coincide with saddle-node bifurcations. Panel (b): Comparison of the available potential energy (APE) of the simulation state (blue line) with the poloidal kinetic energy (PKE; red line). The dashed red line shows the PKE of a simulation with $\gamma = 2$, for which the MHD pedestal is not metastable. Panel (c): Available heat load (AHL) compared with actual thermal drain in the simulation. Quantities in panels (b) and (c) are measured in units of $\Delta E_H$, defined to be the excess thermal energy in H-mode vs. L-mode.}
    \label{fig:cycle}
\end{figure*}

\mbox{\textit{Bifurcation diagram.}\quad} Fig.~\ref{fig:cycle}(a) shows the interchange acceleration [Eq.~\eqref{interchange_force}] of a flux tube moved radially outward from $r_1=\Rh$ at each time in the simulation, where we obtain 1D profiles by averaging $\bQ$ over~$z$. Vertical lines show the times visualized in Fig.~\ref{fig:collapse}. At each time, there are typically three equilibria available to such a tube: $r=1.0$ (which is unstable, because the tube was recently heated), $r\simeq R_{\mathrm{ped}}=1.25 \Rh$ (stable) and another at $r\gtrsim R_{\mathrm{ped}}$ (unstable). We see that bursting events correspond to the collision and annihilation of the latter two equilibria in a saddle-node bifurcation.

\mbox{\textit{Available potential energy.\quad}} To analyze the collective dynamics of the flux tubes over many collapse events, we consider their \textit{available potential energy (APE)}. This is the difference between the total magnetic, thermal and azimuthal kinetic energies, and the \emph{smallest} value that this total can possibly take under any combination of (\emph{i}) nonlinear flux-tube interchanges and (\emph{ii}) ``draining'' of plasma to the initial condition at the outer sink. The concept of APE is well known in geophysical contexts~\cite{Lorenz55}, and was first used to describe plasmas by Ref.~\cite{HoskingCowley24}, for the case of 2D stratified plasma atmospheres under gravity. Calculating APE for the Z-pinch involves solving a nonlinear assignment problem (NAP) for the optimal assignment of flux tubes to different (discretized) radii. In general, the NAP is NP-hard and requires heuristic algorithms. However, in the End Matter we show via an asymptotic expansion that, as $\beta \to 0$, the problem reduces to \emph{linear sum assignment}, which is solvable via the so-called Hungarian or Kuhn--Munkres algorithm~\cite{Kuhn55, Munkres57}. We also present a heuristic method for the NAP at finite $\beta$, which is guaranteed to find equilibria whose energy is increased by any displacement of a small volume of plasma, even though energetic profitability of large-volume displacements cannot be excluded.  

\mbox{\textit{Energetics in the simulation. \quad}} Fig.~\ref{fig:cycle}(b) compares the poloidal kinetic energy (PKE) from the simulation (red line) with the APE. We observe a cycle of $15$ eruption events, each separated by $\Delta {t_{\mathrm{erupt}}\simeq 500 \Rh/c_{s,0}(\Rh)\sim \tau_{\mathrm{heat}}}$. In each case, the APE grows during quiescent periods where the PKE is small, reaching a maximum where instability is triggered---at this point, the APE decays, with accompanying rapid growth of the PKE. The spikes of increased PKE are localized to the time periods in which the APE is decaying, a consequence of dissipation of PKE by poloidal drag [see below Eq.~\eqref{p_H}]. The drag is also the cause of the inefficient transfer of APE to PKE (variations in the APE are around five times larger than those in the PKE). In the Supplementary Information, we plot the cumulative energy dissipated by drag, confirming that it accounts for the deficit. A dashed red line shows the kinetic energy in an identical simulation but with ${\gamma = 2}$, for which the MHD pedestal is not metastable [Eq.~\eqref{interchange_force}]. This simulation does not exhibit periodic bursts, but instead erupts continuously (a movie of that simulation is available at~\url{https://youtu.be/NXJqugt375w}).

Fig.~\ref{fig:cycle}(c) shows the temporal evolution of the heat delivered to the sink near the outer wall (see Fig.~\ref{fig:LH_comparison}). The pale pink line shows the instantaneous thermal-energy drain rate (TDR). We see that periods of increased TDR are centered on times when the PKE peaks. Variations in the TDR occur at a much higher frequency than variations in the PKE; this is because each burst is associated with the expulsion of \emph{many} (typically around $10$) small filaments from the H-mode column. The blue line in panel~(c) shows the available heat load (AHL)---we define this to be the heat that \emph{would} be drained by the sink in forming the minimum-energy configuration identified in the calculation of APE (see above). The AHL reduces by a factor of $\sim 0.5$ during each burst, similar to the factor by which the APE reduces---this indicates that each relaxation is somewhat incomplete. We also plot in red the cumulative thermal drain (CTD; the time-integral of the TDR). The CTD over each burst is around six times the characteristic change in the AHL. The discrepancy can be attributed to the fact that each eruption, while fast, delivers only a small amount of hot plasma to the sink, and many, occurring over a finite fraction of $\tau_{\mathrm{heat}}$, are involved in each collapse event.

\mbox{\textit{Discussion}.\quad} We have shown that a Z-pinch with sheared field-aligned flow can support an internal transport barrier with many similarities to the pedestal region of tokamak fusion devices operating in H-mode. In particular, the Z-pinch ``MHD pedestal'' can be metastable, and we have shown that plasma heating that pushes the pinch to the stability boundary can trigger its collapse by the eruption of fast-moving [$u_r\sim c_s$, see Eq.~\eqref{singlefluxtubemotion}] filaments, analogous to type-I ELMs in tokamaks.

While our simplified model reproduces the above features of ELM bursts in tokamaks, the main qualitative difference is that, in our model, many successive eruptions, over a finite fraction of the heating timescale, appear to be necessary to collapse the pedestal, while in real ELM events, a single eruption event (possibly corresponding to many filaments across the device) appears to be sufficient. An immediate resolution to this discrepancy is that, in our model, flux tubes are always parallel, and therefore the possibility of \emph{magnetic reconnection} creating pathways for fast drainage of hot plasma to the wall along field lines is excluded. A second possibility is that, in real devices, displacing a filament is not energetically profitable unless it has sufficient width to modify the local equilibrium. It is interesting in this context to note that, at $\beta\sim 1$, the assignment problem in a Z-pinch becomes nonlinear and, therefore, the Z-pinch could, in principle, support equilibria for which displacing a small volume of plasma is not energetically profitable, whereas displacing a sufficiently large volume may be.

\textit{Acknowledgments.\quad}We are grateful to S.~C. Cowley and J.~Squire for encouragement and stimulating discussions. It is also a pleasure to acknowledge conversations with T.~Adkins, S.~J.~Benavides, W.~Clarke, T.~Foster, G.~Hammett, M.~W.~Kunz, F.~Marcotte, F.~I.~Parra, E.~Quataert, A.~A.~Schekochihin, C.~Skene, A.~Spitkovsky, S.~Tobias and H.~R.~Wilson. LS and RK are grateful to OpenStar Technologies for supporting their work on this project as part of the University of Cambridge's \emph{Summer Research in Mathematics (SRIM)} program.
\vspace{1mm}

We are grateful to have known our friend N.~F.~G.~Loureiro, with whom it was a pleasure to discuss this and many other projects.

\bibliography{decay_mod_prx.bib}

@ARTICLE{Kuhn55,
       author = {{Kuhn}, H.~W.},
        title = {{The Hungarian method for the assignment problem}},
      journal = {Nav. Res. Logist. Quart.},
       volume = {2},
        pages = {83},
          doi = {https://doi.org/10.1002/nav.3800020109},
          url = {https://onlinelibrary.wiley.com/doi/abs/10.1002/nav.3800020109},
         year = {1955}
}

@ARTICLE{Munkres57,
         ISSN = {03684245},
          URL = {http://www.jstor.org/stable/2098689},
       author = {{Munkres}, J},
      journal = {J. Soc. Ind. Appl. Math.},
        pages = {32},
    publisher = {Society for Industrial and Applied Mathematics},
        title = {Algorithms for the assignment and transportation problems},
      urldate = {2023-09-15},
       volume = {5},
         year = {1957}
}

@ARTICLE{CowleyArtun97,
       author = {{Cowley}, S.~C. and {Artun}, M.},
        title = "{Explosive instabilities and detonation in magnetohydrodynamics.}",
      journal = {Phys. Rep.},
         year = 1997,
        month = apr,
       volume = {283},
        pages = {185},
          doi = {10.1016/S0370-1573(96)00060-9},
       adsurl = {https://ui.adsabs.harvard.edu/abs/1997PhR...283..185C}
}

@ARTICLE{CowleyCowley15,
       author = {{Cowley}, S.~C. and {Cowley}, B. and {Henneberg}, S.~A. and {Wilson}, H.~R.},
        title = "{Explosive instability and erupting flux tubes in a magnetized plasma}",
      journal = {Proc. R. Soc. Lond.},
         year = 2015,
        month = aug,
       volume = {471},
        pages = {20140913},
          doi = {10.1098/rspa.2014.0913},
 primaryClass = {physics.plasm-ph},
       adsurl = {https://ui.adsabs.harvard.edu/abs/2015RSPSA.47140913C}
}

@ARTICLE{ITER99_c2,
       author = {{ITER Physics Expert Group on Confinement and Transport} and {ITER Physics Expert Group on Confinement Modelling} and {ITER Physics Basis Editors}},
        title = "{ITER physics basis, chapter 2: Plasma confinement and transport}",
      journal = {Nucl. Fusion},
         year = 1999,
        month = dec,
       volume = {39},
        pages = {2175},
          doi = {10.1088/0029-5515/39/12/302},
       adsurl = {https://ui.adsabs.harvard.edu/abs/1999NucFu..39.2175I}
}

@ARTICLE{ITER07_c3,
       author = {{Hender}, T.~C. and {Wesley}, J.~C. and {Bialek}, J. and {Bondeson}, A. and {Boozer}, A.~H. and {Buttery}, R.~J. and {Garofalo}, A. and {Goodman}, T.~P. and {Granetz}, R.~S. and {Gribov}, Y. and others},
        title = "{Progress in the ITER physics basis, chapter 3: MHD stability, operational limits and disruptions}",
      journal = {Nucl. Fusion},
         year = 2007,
        month = jun,
       volume = {47},
        pages = {S128},
          doi = {10.1088/0029-5515/47/6/S03},
       adsurl = {https://ui.adsabs.harvard.edu/abs/2007NucFu..47S.128H}
}

@ARTICLE{Lorenz55,
       author = {{Lorenz}, E.~N.},
        title = "{Available potential energy and the maintenance of the general circulation}",
      journal = {Tellus},
         year = 1955,
        month = jan,
       volume = {7},
        pages = {157},
          doi = {10.3402/tellusa.v7i2.8796},
       adsurl = {https://ui.adsabs.harvard.edu/abs/1955Tell....7..157L}
}

@ARTICLE{HoskingCowley24,
       author = {{Hosking}, D.~N. and {Wasserman}, D. and {Cowley}, S.~C.},
        title = "{Metastability of stratified magnetohydrostatic equilibria and their relaxation}",
      journal = {J. Plasma Phys.},
         year = 2025,
        month = feb,
       volume = {91},
          eid = {E35},
        pages = {E35},
          doi = {10.1017/S0022377824001521},
 primaryClass = {astro-ph.HE},
       adsurl = {https://ui.adsabs.harvard.edu/abs/2025JPlPh..91E..35H}
}

@ARTICLE{Helander99,
       author = {{Helander}, P. and {Chapman}, S.~C. and {Dendy}, R.~O. and {Rowlands}, G. and {Watkins}, N.~W.},
        title = "{Exactly solvable sandpile with fractal avalanching}",
      journal = {Phys. Rev. E},
         year = 1999,
        month = jun,
       volume = {59},
        pages = {6356},
          doi = {10.1103/PhysRevE.59.6356},
       adsurl = {https://ui.adsabs.harvard.edu/abs/1999PhRvE..59.6356H}
}

@ARTICLE{Hughes20_SPARC,
       author = {{Hughes}, J.~W. and {Howard}, N.~T. and {Rodriguez-Fernandez}, P. and {Creely}, A.~J. and {Kuang}, A.~Q. and {Snyder}, P.~B. and {Wilks}, T.~M. and {Sweeney}, R. and {Greenwald}, M.},
        title = "{Projections of H-mode access and edge pedestal in the SPARC tokamak}",
      journal = {J. Plasma Phys.},
         year = 2020,
        month = oct,
       volume = {86},
          eid = {865860504},
        pages = {865860504},
          doi = {10.1017/S0022377820001300},
       adsurl = {https://ui.adsabs.harvard.edu/abs/2020JPlPh..86e8604H}
}

@ARTICLE{GuazzottoBetti11,
       author = {{Guazzotto}, L. and {Betti}, R.},
        title = "{Magnetohydrodynamic mechanism for pedestal formation}",
      journal = {Phys. Rev. Lett.},
         year = 2011,
        month = sep,
       volume = {107},
          eid = {125002},
        pages = {125002},
          doi = {10.1103/PhysRevLett.107.125002},
       adsurl = {https://ui.adsabs.harvard.edu/abs/2011PhRvL.107l5002G}
}

@ARTICLE{Lennholm24_STEP,
       author = {{Lennholm}, M. and {Aleiferis}, S. and {Bakes}, S. and {Bardsley}, O.~P. and {van Berkel}, M. and {Casson}, F.~J. and {Chaudry}, F. and {Conway}, N.~J. and {Hender}, T.~C. and {Henderson}, S.~S. and others},
        title = "{Plasma control for the step prototype power plant}",
      journal = {Nucl. Fusion},
         year = 2024,
        month = sep,
       volume = {64},
          eid = {096036},
        pages = {096036},
          doi = {10.1088/1741-4326/ad6012},
       adsurl = {https://ui.adsabs.harvard.edu/abs/2024NucFu..64i6036L}
}

@ARTICLE{Tholerus24_STEP,
       author = {{Tholerus}, E. and {Casson}, F.~J. and {Marsden}, S.~P. and {Wilson}, T. and {Brunetti}, D. and {Fox}, P. and {Freethy}, S.~J. and {Hender}, T.~C. and {Henderson}, S.~S. and {Hudoba}, A. and others},
        title = "{Flat-top plasma operational space of the STEP power plant}",
      journal = {Nucl. Fusion},
         year = 2024,
        month = oct,
       volume = {64},
          eid = {106030},
        pages = {106030},
          doi = {10.1088/1741-4326/ad6ea2},
 primaryClass = {physics.plasm-ph},
       adsurl = {https://ui.adsabs.harvard.edu/abs/2024NucFu..64j6030T}
}

@ARTICLE{Wagner82_ASDEX,
       author = {{Wagner}, F. and {Becker}, G. and {Behringer}, K. and {Campbell}, D. and {Eberhagen}, A. and {Engelhardt}, W. and {Fussmann}, G. and {Gehre}, O. and {Gernhardt}, J. and {Gierke}, G.~V. and others},
        title = "{Regime of improved confinement and high beta in neutral-beam-heated divertor discharges of the ASDEX tokamak}",
      journal = {Phys. Rev. Lett.},
         year = 1982,
        month = nov,
       volume = {49},
        pages = {1408},
          doi = {10.1103/PhysRevLett.49.1408},
       adsurl = {https://ui.adsabs.harvard.edu/abs/1982PhRvL..49.1408W}
}

@ARTICLE{Wagner84_ASDEX,
       author = {{Wagner}, F. and {Fussmann}, G. and {Grave}, T. and {Keilhacker}, M. and {Kornherr}, M. and {Lackner}, K. and {McCormick}, K. and {M{\"u}ller}, E.~R. and {St{\"a}bler}, A. and {Becker}, G. and others},
        title = "{Development of an edge transport barrier at the H-mode transition of ASDEX}",
      journal = {Phys. Rev. Lett.},
         year = 1984,
        month = oct,
       volume = {53},
        pages = {1453},
          doi = {10.1103/PhysRevLett.53.1453},
       adsurl = {https://ui.adsabs.harvard.edu/abs/1984PhRvL..53.1453W}
}

@ARTICLE{Zohm96_ELMs,
       author = {{Zohm}, H.},
        title = "{Edge localized modes (ELMs)}",
      journal = {Plasma Phys. Control. Fusion},
         year = 1996,
        month = feb,
       volume = {38},
        pages = {105},
          doi = {10.1088/0741-3335/38/2/001},
       adsurl = {https://ui.adsabs.harvard.edu/abs/1996PPCF...38..105Z}
}

@ARTICLE{Gimblett06_ELMs,
       author = {{Gimblett}, C.~G. and {Hastie}, R.~J. and {Helander}, P.},
        title = "{Model for current-driven edge-localized modes}",
      journal = {Phys. Rev. Lett.},
         year = 2006,
        month = jan,
       volume = {96},
          eid = {035006},
        pages = {035006},
          doi = {10.1103/PhysRevLett.96.035006},
       adsurl = {https://ui.adsabs.harvard.edu/abs/2006PhRvL..96c5006G}
}

@ARTICLE{Chapman01_sandpile,
       author = {{Chapman}, S.~C. and {Dendy}, R.~O. and {Hnat}, B.},
        title = "{Sandpile model with tokamak-like enhanced confinement phenomenology}",
      journal = {Phys. Rev. Lett.},
         year = 2001,
        month = mar,
       volume = {86},
        pages = {2814},
          doi = {10.1103/PhysRevLett.86.2814},
 primaryClass = {physics.plasm-ph},
       adsurl = {https://ui.adsabs.harvard.edu/abs/2001PhRvL..86.2814C}
}

@ARTICLE{Chapman01b_sandpile,
       author = {{Chapman}, S.~C. and {Dendy}, R.~O. and {Rowlands}, G.},
        title = "{A sandpile model with dual scaling regimes for laboratory, space and astrophysical plasmas}",
      journal = {Phys. Plasmas},
         year = 1999,
        month = nov,
       volume = {6},
        pages = {4169},
          doi = {10.1063/1.873682},
       adsurl = {https://ui.adsabs.harvard.edu/abs/1999PhPl....6.4169C}
}

@INPROCEEDINGS{ConnorKirkWilson08_ELMs,
       author = {{Connor}, J.~W. and {Kirk}, A. and {Wilson}, H.~R.},
        title = "{Edge localised modes (ELMs): Experiments and theory}",
    booktitle = {Turbulent Transport in Fusion Plasmas},
         year = 2008,
       editor = {{Benkadda}, Sadruddin},
       series = {American Institute of Physics Conference Series},
       volume = {1013},
        month = may,
    publisher = {AIP},
        pages = {174},
          doi = {10.1063/1.2939030},
       adsurl = {https://ui.adsabs.harvard.edu/abs/2008AIPC.1013..174C}
}

@ARTICLE{Cathey20_Jorek,
       author = {{Cathey}, A. and {Hoelzl}, M. and {Lackner}, K. and {Huijsmans}, G.~T.~A. and {Dunne}, M.~G. and {Wolfrum}, E. and {Pamela}, S.~J.~P. and {Orain}, F. and {G{\"u}nter}, S. and {JOREK Team} and others},
        title = "{Non-linear extended MHD simulations of type-I edge localised mode cycles in ASDEX Upgrade and their underlying triggering mechanism}",
      journal = {Nucl. Fusion},
         year = 2020,
        month = dec,
       volume = {60},
          eid = {124007},
        pages = {124007},
          doi = {10.1088/1741-4326/abbc87},
 primaryClass = {physics.plasm-ph},
       adsurl = {https://ui.adsabs.harvard.edu/abs/2020NucFu..60l4007C}
}

@ARTICLE{Stone20,
       author = {{Stone}, J.~M. and {Tomida}, K. and {White}, C.~J. and {Felker}, K.~G.},
        title = "{The Athena++ adaptive mesh refinement framework: Design and magnetohydrodynamic solvers}",
      journal = {Astrophys. J. Suppl.},
         year = 2020,
        month = jul,
       volume = {249},
          eid = {4},
        pages = {4},
          doi = {10.3847/1538-4365/ab929b},
 primaryClass = {astro-ph.IM},
       adsurl = {https://ui.adsabs.harvard.edu/abs/2020ApJS..249....4S}
}

@ARTICLE{Snyder11_EPED,
       author = {{Snyder}, P.~B. and {Groebner}, R.~J. and {Hughes}, J.~W. and {Osborne}, T.~H. and {Beurskens}, M. and {Leonard}, A.~W. and {Wilson}, H.~R. and {Xu}, X.~Q.},
        title = "{A first-principles predictive model of the pedestal height and width: development, testing and ITER optimization with the EPED model}",
      journal = {Nucl. Fusion},
         year = 2011,
        month = oct,
       volume = {51},
          eid = {103016},
        pages = {103016},
          doi = {10.1088/0029-5515/51/10/103016},
       adsurl = {https://ui.adsabs.harvard.edu/abs/2011NucFu..51j3016S}
}

@ARTICLE{Snyder12_EPED,
       author = {{Snyder}, P.~B. and {Osborne}, T.~H. and {Burrell}, K.~H. and {Groebner}, R.~J. and {Leonard}, A.~W. and {Nazikian}, R. and {Orlov}, D.~M. and {Schmitz}, O. and {Wade}, M.~R. and {Wilson}, H.~R.},
        title = "{The EPED pedestal model and edge localized mode-suppressed regimes: Studies of quiescent H-mode and development of a model for edge localized mode suppression via resonant magnetic perturbations}",
      journal = {Phys. Plasmas},
         year = 2012,
        month = may,
       volume = {19},
          eid = {056115},
        pages = {056115},
          doi = {10.1063/1.3699623},
       adsurl = {https://ui.adsabs.harvard.edu/abs/2012PhPl...19e6115S}
}

@ARTICLE{Groebner13_EPED,
       author = {{Groebner}, R.~J. and {Chang}, C.~S. and {Hughes}, J.~W. and {Maingi}, R. and {Snyder}, P.~B. and {Xu}, X.~Q. and {Boedo}, J.~A. and {Boyle}, D.~P. and {Callen}, J.~D. and {Canik}, J.~M. and others},
        title = "{Improved understanding of physics processes in pedestal structure, leading to improved predictive capability for ITER}",
      journal = {Nucl. Fusion},
         year = 2013,
        month = sep,
       volume = {53},
          eid = {093024},
        pages = {093024},
          doi = {10.1088/0029-5515/53/9/093024},
       adsurl = {https://ui.adsabs.harvard.edu/abs/2013NucFu..53i3024G}
}

@ARTICLE{Walk12,
       author = {{Walk}, J.~R. and {Snyder}, P.~B. and {Hughes}, J.~W. and {Terry}, J.~L. and {Hubbard}, A.~E. and {Phillips}, P.~E.},
        title = "{Characterization of the pedestal in Alcator C-Mod ELMing H-modes and comparison with the EPED model}",
      journal = {Nucl. Fusion},
         year = 2012,
        month = jun,
       volume = {52},
          eid = {063011},
        pages = {063011},
          doi = {10.1088/0029-5515/52/6/063011},
       adsurl = {https://ui.adsabs.harvard.edu/abs/2012NucFu..52f3011W}
}

@ARTICLE{Garbet04_stiffness,
       author = {{Garbet}, X. and {Mantica}, P. and {Ryter}, F. and {Cordey}, G. and {Imbeaux}, F. and {Sozzi}, C. and {Manini}, A. and {Asp}, E. and {Parail}, V. and {Wolf}, R. and others},
        title = "{Profile stiffness and global confinement}",
      journal = {Plasma Phys. Control. Fusion},
         year = 2004,
        month = sep,
       volume = {46},
        pages = {1351},
          doi = {10.1088/0741-3335/46/9/002},
       adsurl = {https://ui.adsabs.harvard.edu/abs/2004PPCF...46.1351G}
}

@ARTICLE{Ryter01_stiffness,
       author = {{Ryter}, F. and {Stober}, J. and {St{\"a}bler}, A. and {Tardini}, G. and {Fahrbach}, H.-U. and {Gruber}, O. and {Herrmann}, A. and {Kallenbach}, A. and {Kaufmann}, M. and {Kurzan}, B. and others},
        title = "{Confinement and transport studies of conventional scenarios in ASDEX Upgrade}",
      journal = {Nucl. Fusion},
         year = 2001,
        month = may,
       volume = {41},
        pages = {537},
          doi = {10.1088/0029-5515/41/5/307},
       adsurl = {https://ui.adsabs.harvard.edu/abs/2001NucFu..41..537R}
}

\newpage
\newpage
\clearpage

\setcounter{page}{1}

\section*{End Matter}

\mbox{\textit{Interchange stability to $\mathcal{O}(\delta r^2)$}.\quad} 
Expanding Eq.~\eqref{interchange_force} in small $\delta r=r_2-r_1$, we have
\begin{equation}
    \frac{\dd u_r}{\dd t} = a_{1\to2}= -\mathcal{L}(r_1)\delta r + \mathcal{N}(r_1)\delta r^2 + \mathcal{O}(\delta r^3)
\end{equation}where
\begin{equation}
    \mathcal{L}(r) = \frac{2\chi I^{\gamma-1}}{r^{2\gamma-1}}\Sigma'(r)+ \frac{\chi}{r^3} \Lambda'(r)\label{Lapp}
\end{equation}and
\begin{multline}
    \mathcal{N}(r)= \frac{2\chi I^{\gamma-1}}{r^{2\gamma-1}}\left[\frac{2\gamma-1}{r}\Sigma'(r)-\frac{1}{2}\Sigma''(r)\right]\\
    +\frac{\chi}{r^3}\left[\frac{3}{r}\Lambda'(r)-\frac{1}{2}\Lambda''(r)\right],
\end{multline}where $\Sigma = s^{\gamma}/\chi^{\gamma}$, $\Lambda = \ell^2/\chi$ and primes denote derivatives with respect to $r$.

At marginal linear stability, $\mathcal{L}=0$ for all $r$. Setting Eq.~\eqref{Lapp} and its first derivative with respect to $r$ to zero, we find
\begin{align}
    \Lambda'(r) & = - 2I^{\gamma-1} r^{4-2\gamma}\Sigma'(r)\label{lambdaprimeapp},\\
    \Lambda''(r) & = - 2I^{\gamma-1}r^{3-2\gamma}\left[(4-2\gamma)\Sigma'(r)+r\Sigma''(r)\right],
\end{align}from which it follows that
\begin{equation}
    \mathcal{N}(r)=-2(2-\gamma)\frac{I^{\gamma-1}}{r^{2\gamma}}\Sigma'(r).
\end{equation}Thus, at marginal linear stability, there remains a nonlinear force in the direction of decreasing $\Sigma$. The presence of such a force requires the Z-pinch to be rotating, as Eq.~\eqref{lambdaprimeapp} yields $\Sigma'=0$ if $\Lambda(r)=0$ for all~$r$.

\mbox{\textit{Available potential energy (APE).\quad}} We formulate the nonlinear assignment problem (NAP) outlined in the main text as follows. Let $\bQ_{\mathrm{id}}(\psi)$ define a 1D profile of Lagrangian invariants in the flux coordinate
\begin{equation}
    \psi=\int_{r_{\mathrm{min}}}^r \dd r \, B(r)
\end{equation}that is piecewise constant in equally spaced flux bins of width $\Delta \psi$---we explain in the Supplementary Information how we obtain $\bQ_{\mathrm{id}}(\psi)$ from 2D simulation states. From this base profile, we define $\bQ_\sigma$ to be the profile obtained by permuting $\bQ_{\mathrm{id}}(\psi)$ over the flux bins, with permutation $\sigma$. We define $P_{\sigma}(\psi)$ and $r_{\sigma}(\psi)$ to be the \emph{continuous} equilibrium pressure and radius, respectively, corresponding to $\bQ_{\sigma}$. We obtain them by integrating simultaneously
\begin{equation}
    \frac{\dd P_{\sigma}}{\dd \psi} = -\frac{f_{\sigma}}{B_{\sigma}}, \quad \frac{\dd r_\sigma}{\dd \psi} = \frac{1}{B_{\sigma}}\label{Psigma_rsigma}
\end{equation}where $f_{\sigma}\equiv f(P_{\sigma},r_\sigma, \bQ_{\sigma})$ and $B_{\sigma}\equiv B(P_{\sigma},r_{\sigma}, \bQ_{\sigma})$. We choose, without loss of generality, $\psi = 0$ at $r=r_{\mathrm{min}}$, so integration constants are fixed by $r_{\sigma}(0)=r_{\mathrm{min}}$ and $r_{\sigma}(\psi_\mathrm{max})=r_{\mathrm{max}}$.  The energy $E_{\sigma}$ associated with a given permutation~$\sigma$ is
\begin{align}
     E_{\sigma} = \int \dd \psi \,\mcE_{\sigma},
\end{align}where
\begin{equation}
    \mcE_\sigma \equiv \frac{1}{\rho_{\sigma} \chi_{\sigma}}\left(\frac{\rho_{\sigma}^{\gamma}s_{\sigma}^{\gamma}}{\gamma - 1}+\frac{\rho_{\sigma}^2 r_{\sigma}^2 \chi_{\sigma}^2}{2}+\frac{\rho_{\sigma}\ell_{\sigma}^2}{2r_{\sigma}^2}\right).\label{E}
\end{equation}and $\rho_{\sigma}\equiv\rho(P_{\sigma},r_\sigma, \bQ_{\sigma})$. The APE is the solution of the NAP defined by the minimization of $E_{\sigma}$ over assignments~$\sigma$.

\mbox{\textit{Case of $\beta\to 0$: \quad}} In the case of interest for this paper, viz., $\beta\to 0$ with $\Ma\sim 1$, the NAP detailed above reduces to a linear sum assignment problem (LSAP). To show this, we write 
\begin{equation}
    \bQ_{\sigma} = (s_{\sigma},\chi_{\sigma},\ell_{\sigma}) = (\epsilon^{1/\gamma}S_{\sigma}, C_{\sigma}, \epsilon^{1/2}L_{\sigma}) \label{Qepsilon}
\end{equation}and form an asymptotic expansion of Eq.~\eqref{E} in $\epsilon\ll 1$ with $S_{\sigma}$, $C_{\sigma}$, and $L_{\sigma}$ fixed. The exponents of $\epsilon$ in Eq.~\eqref{Qepsilon} ensure that the expansion is in integer powers of $\epsilon$ and that $\Ma \sim 1$. In what follows, we express the expansions of $P_{\sigma}$, $r_{\sigma}$, $\mcE_\sigma$, $E_\sigma$ and $\rho_\sigma$ in the form ${X = X_0+\epsilon X_1 + \mathcal{O}(\epsilon^2)}$, and suppress $\sigma$ subscripts for clarity.

\mbox{\textit{Expansion at $\mathcal{O}(\epsilon^0)$:\quad }} At leading order in $\epsilon$, it is straightforward to show that
\begin{equation}
    P_0(\psi) = \frac{I^2}{2r_0(\psi)^2},\quad\rho_0(\psi)=\frac{I^2}{Cr_0(\psi)^2},\label{P0rho0}
\end{equation}where
\begin{equation}
    r_0(\psi) = r_{\mathrm{min}}\exp\left(\frac{\psi}{I}\right), \quad I = \frac{\psi_{\mathrm{max}}}{\ln(r_{\mathrm{max}}/r_{\mathrm{min}})}.\label{r0ofpsi}
\end{equation}The constant $I$ is readily interpreted as the current through the central wire. Eqs.~\eqref{P0rho0} and~\eqref{r0ofpsi} yield
\begin{equation}
    \mathcal{E}_0 = \frac{1}{2}\rho_0C r_0^2,\quad E_0 = \frac{1}{2}I^2\ln \frac{r_{\mathrm{max}}}{r_{\mathrm{min}}},
\end{equation}so the leading order energy (being the energy associated with the magnetic field in vacuum) is independent of the permutation $\sigma$. Breaking the degeneracy requires the $\mathcal{O}(\epsilon)$ terms in the expansion.

\mbox{\textit{Expansion at  $\mathcal{O}(\epsilon^1)$:\quad }}Using $s^{\gamma}=\epsilon S^{\gamma}$ and $\ell^2 = \epsilon L^2$, we have
\begin{equation}
\mathcal{E}_1=\frac{1}{\gamma - 1}\frac{\rho_0^{\gamma-1}S^{\gamma}}{C}+\frac{1}{2}C(\rho_1r_0^2+2\rho_0r_0r_1)+\frac{L^2}{2Cr_0^2}
.\end{equation}To find $\rho_1$, we use the exact relation $\dd r/\dd\psi = 1/\rho C r$ [Eq.~\eqref{Psigma_rsigma}] to yield $\rho_1 = -\rho_0[r_1/r_0+(\dd r_1/\dd \psi)/(\dd r_0/\dd \psi)]$. Some manipulation using Eq.~\eqref{r0ofpsi} yields
\begin{equation}
     \mathcal{E}_1=\frac{1}{\gamma - 1}\frac{\rho_0^{\gamma-1}S^{\gamma}}{C}+\frac{L^2}{2Cr_0^2}-\frac{1}{2}\frac{\dd}{\dd \psi}\left[r_0 r_1\left(\frac{\dd r_0}{\dd \psi}\right)^{-1}\right],\label{mcE1}
\end{equation}where $r_1(\psi)$ is determined by integrating the first of Eqs.~\eqref{Psigma_rsigma}. However, the boundary conditions $r_{\sigma}(0)=r_{\mathrm{min}}$ and $r_{\sigma}(\psi_{\mathrm{max}})=r_{\mathrm{max}}$ require that $r_1(0)=r_1(\psi_{\mathrm{max}})=0$, so the term involving $r_1$ in Eq.~\eqref{mcE1} vanishes after integration in $\psi$. Thus, up to $\mathcal{O}(\epsilon^2)$,
\begin{equation}
    E_{\sigma} = E_0 + \int_0^{\psi_{\mathrm{max}}} \dd \psi \Bigg[\frac{K}{r_0(\psi)^{2\gamma-2}}\left(\frac{s_{\sigma}}{\chi_{\sigma}}\right)^{\gamma}+\frac{1}{2 r_0(\psi)^2}\frac{\ell_{\sigma}^2}{\chi_{\sigma} }\Bigg],\label{lowbeta_LSAP}
\end{equation}where $K=I^{\gamma-1}/(\gamma-1)$. Because the integrand appearing in Eq.~\eqref{lowbeta_LSAP} is a \emph{function}, rather than a \emph{functional} of $\psi$ and $\bQ_{\sigma}$, minimizing $E_{\sigma}$ to first order in $\epsilon$ is an LSAP.

\mbox{\textit{Solution for $\beta\sim 1$:\quad}}  A physically motivated heuristic procedure to solve the NAP is as follows. Consider the simpler assignment problem of, given a particular stepped equilibrium $\bQ_{\sigma}(\psi)$, finding the optimal assignment $\pi$ of equal infinitesimal magnetic flux tubes from each step to other steps. The MHD equations for a Z-pinch (see Supplementary Information) imply that
\begin{equation}
    \frac{\dd}{\dd t}\int_{\mathcal{M}(t)} \dd V\left (e + P + \frac{\rho u_{\mathrm{pol}}^2}{2}\right) = \int_{\mathcal{M}(t)} \dd V\, \frac{\p P}{\p t},
\end{equation}where $\mathcal{M}(t)$ is a material volume and $e=\rho \chi \mcE$ is the internal energy density. If pressure is everywhere constant in an Eulerian sense because the volume of moved fluid is small, the right-hand side vanishes, showing that there is a fixed budget for the sum of enthalpy and poloidal kinetic energy. The most profitable assignment is the one that minimizes
\begin{equation}
     H_{\sigma\pi}\equiv \int \dd \psi \,h(P_{\sigma}, r_{\sigma}, \bQ_{\pi}),\quad h \equiv \frac{e+P}{\rho \chi}\label{h}
\end{equation}over $\pi$, where $h$ is enthalpy per unit flux. This is a linear assignment problem because $\sigma$ is fixed while $\pi$ is optimized. The solution for $\pi$ can be viewed as a candidate solution for $\sigma$: setting $\sigma=\pi$, one may solve Eqs.~\eqref{Psigma_rsigma} for a new equilibrium. Iterating this procedure to convergence yields a final equilibrium that, while not guaranteed to be globally optimal, cannot be improved by nonlinear displacement of small volumes of plasma (by construction). To demonstrate that, at low $\beta$, this approach is consistent with Eq.~\eqref{lowbeta_LSAP}, we note that a similar asymptotic calculation to the one in the previous section yields
\begin{multline}
    H_{\sigma\pi} = 2E_0 + \epsilon \int \dd \psi \left[r_{\sigma,1}\sqrt{2P_{\sigma,0}}+\frac{r_{\sigma,0} P_{\sigma,1}}{\sqrt{2P_{\sigma,0}}} \right] \\ +\int_0^{\psi_{\mathrm{max}}} \dd \psi \Bigg[\frac{K}{r_0(\psi)^{2\gamma-2}}\left(\frac{s_{\pi}}{\chi_{\pi}}\right)^{\gamma}+\frac{1}{2 r_0(\psi)^2}\frac{\ell_{\pi}^2}{\chi_{\pi} }\Bigg] + \mathcal{O}(\epsilon^2).\label{enthalpy_minimization_EndMatter}
\end{multline}Thus, the minimization problem for $\pi$ at fixed $\sigma$ at $\beta \ll 1$ is precisely the same as the one defined by Eq.~\eqref{lowbeta_LSAP}. 

\mbox{\textit{Case of a sink.} \quad} We now generalize the low-$\beta$ LSAP [Eq.~\eqref{lowbeta_LSAP}] to include a fast sink that can reset the invariants of a flux tube when it reaches the wall and drains, as in our simulations (the iterative procedure for the $\beta\sim 1$ NAP can be generalized along similar lines). 

Plasma with invariants $\bQ_j$ can contribute to the energy at a candidate destination shell $\psi_i$ in two ways: either it remains undrained and carries $\bQ_j$ to shell $i$, or it drains at the sink (which we take to be located at $\psi_{\mathrm{max}}$ for simplicity) and re-enters the system with reference invariants $\bQ_{\mathrm{ref}}$. The respective contributions to the total energy stored by the pinch are
\begin{equation}
    C^{\mathrm{undr}}_{ij} = C(\psi_i, \bQ_j) = \frac{K}{r_0(\psi_i)^{2\gamma-2}}\left(\frac{s_{j}}{\chi_{j}}\right)^{\gamma}+\frac{1}{2 r_0(\psi_i)^2}\frac{\ell_{j}^2}{\chi_{j} }
\end{equation}in the undrained case and
\begin{equation}
    C^{\mathrm{dr}}_{i} = C(\psi_i, \bQ_{\mathrm{ref}})
\end{equation}in the drained case. In the case where the plasma is drained at the sink, the \emph{environment} gains energy 
\begin{multline}
    \Delta E^{\mathrm{dr}}_{j} = C(\psi_{\mathrm{max}}, \bQ_{j}) - C(\psi_{\mathrm{max}}, \bQ_{\mathrm{ref}}) \\ = \frac{K}{r_{\mathrm{max}}^{2\gamma -2}}\left[\left(\frac{s_j}{\chi_j}\right)^{\gamma}-\left(\frac{s_{\mathrm{ref}}}{\chi_{\mathrm{ref}}}\right)^{\gamma}\right] + \frac{1}{2r_{\mathrm{max}}^{2}}\left(\frac{\ell_j^2}{\chi_j}-\frac{\ell_{\mathrm{ref}}^2}{\chi_{\mathrm{ref}}}\right),\label{DeltaEdrj}
\end{multline}where we have assumed that the sink is sufficiently slow that negligible energy goes into exciting compressional waves. The available potential energy of the pinch is the maximum energy that can be delivered to motions---since total energy is conserved, this can be determined by minimizing the total energy stored by the pinch and delivered to the environment. This is the state that solves the LSAP
\begin{equation}
    \sum_i \min\left\{C^{\mathrm{undr}}_{i,\sigma(i)},\, C^{\mathrm{dr}}_{i}+\Delta E^{\mathrm{dr}}_{\sigma(i)}\right\}\to \min.\label{leakyLSAP}
\end{equation}
We define the \emph{available heat load (AHL)} to be the thermal part of the total energy transferred to the environment in rearrangement to the most energetically favorable state. From~\eqref{DeltaEdrj}, it is given by
\begin{equation}
    \mathrm{AHL}=  \Delta \psi\sum_{i\in \mathcal{D}} \frac{K}{r_{\mathrm{max}}^{2\gamma -2}}\left[\left(\frac{s_i}{\chi_i}\right)^{\gamma}-\left(\frac{s_{\mathrm{ref}}}{\chi_{\mathrm{ref}}}\right)^{\gamma}\right],\label{AHL}
\end{equation}where $\mathcal{D}$ is the set of drained parcels in the solution to the LSAP~\eqref{leakyLSAP}.

\newpage
\newpage
\clearpage

\setcounter{page}{1}

\begin{center}
    \large \textbf{SUPPLEMENTARY INFORMATION} \normalsize
\end{center}

\section{Assignment problem with 2D input}

\begin{figure*}
    \centering
    \includegraphics[width=\textwidth]{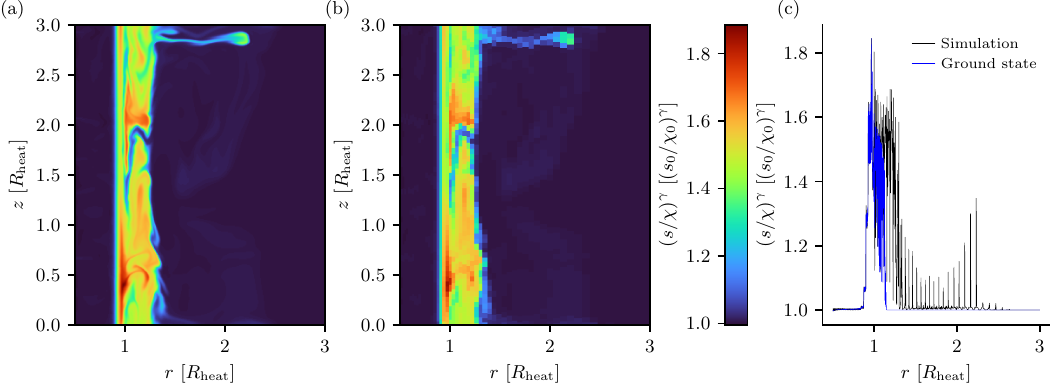}
    \caption{Illustration of the procedure used to obtain a 1D state for use in solution of the assignment problem, as explained in the text. Panel (a) shows the simulation state at a particular instant in time [the same one as in the final panel of Fig.~\ref{fig:collapse}(a)]. Panel (b) shows the same state downsampled into $3000$ equal-flux macrocells over $N_{\mathrm{stripes}}=55$ radial stripes, as described in the text. Panel (c) shows the 1D representation of the simulation state (black line) and the corresponding ground state found by solving the LSAP (blue line).}
    \label{fig:Hungarian_illustration}
\end{figure*}

\mbox{\textit{Numerical setup with macrocells for 2D input.} \quad} The Hungarian algorithm operates on a 1D list of equal-flux units with invariants $\bQ$. Our simulations, however, are 2D, with variables $\rho$, $p$, $B$, and $u_\phi$. We construct a 1D representation of the equilibrium state that can be used in a global assignment problem as follows (a visualization is given in Fig.~\ref{fig:Hungarian_illustration}).

\mbox{\textit{Step 1: Invariants.}} From the simulation output we form the Lagrangian invariants
\begin{align}
    s(r,z)&=\frac{p(r,z)^{1/\gamma}}{\rho(r,z)},\\
    \chi(r,z)&=\frac{B(r,z)}{\rho(r,z)\,r},\\
    \ell(r,z)&=u_\phi(r,z)\,r,
\end{align}
which are the quantities permuted by the algorithm.

\mbox{\textit{Step 2: Radial stripes.}} We partition the radial grid into $N_{\mathrm{stripes}}$ contiguous stripes, each consisting of an integer number of radial cells ($N_{\mathrm{stripes}}=55$ for the calculations presented in the main text). Stripe boundaries are chosen by selecting target radii that are uniformly spaced in $\log r$, and then snapping each target to the nearest available radial cell edge. 

\mbox{\textit{Step 3: 1D flux coordinate.}} Within each stripe $k$ and at each axial location $z$, we define the flux per unit $z$-length as
\begin{equation}
    \mu_k(z) \equiv \frac{1}{L_z}\int_{r\in k}B(r,z)\,\mathrm{d}r,
\end{equation}
where $L_z$ is the axial box size. In practice the $r$-integral is a sum over radial cells in the stripe. We define the cumulative flux within each stripe by
\begin{equation}
    \Psi_k(z)\equiv \int_{z_{\min}}^{z}\mu_k(z')\,\mathrm{d}z',
\end{equation}
using piecewise-linear inversion in $z$ so that boundaries can cut through grid cells.
Finally, we concatenate stripes in order of increasing radius to define a global 1D flux coordinate,
\begin{equation}
    \psi(k,z) \equiv \sum_{k'<k} \Psi_{k'}(z_{\max}) + \Psi_k(z),\quad 0\le \psi\le \psi_{\mathrm{tot}},
\end{equation}
where $\psi_{\mathrm{tot}}=\sum_k \Psi_k(z_{\max})$.

\mbox{\textit{Step 4: Equal-flux macrocells.}} Given a chosen number of flux bins, we define evenly spaced flux edges
\begin{equation}
    \psi_i = \frac{i}{N}\psi_{\mathrm{tot}},\quad i=0,\dots,N,
\end{equation}
so that there are $N$ macrocells of equal flux $\Delta \psi=\psi_{\mathrm{tot}}/N$, where $N=3001$ for the calculations presented in the main text. Each edge $\psi_i$ is mapped to a stripe index $k_i$ and axial position $z_i$ by inverting $\Psi_{k_i}(z_i)=\psi_i-\sum_{k<k_i}\Psi_k(z_{\max})$. A single macrocell $[\psi_i,\psi_{i+1}]$ is therefore represented as a union of one or more $(k,z)$-segments, and can straddle stripe boundaries.

\mbox{\textit{Step 5: Macrocell invariants.}} For each field ${q\in\{s,\chi,\ell\}}$, which are defined \emph{per unit mass}, we compute its macrocell value by taking a mass-weighted average:
\begin{equation}
    q_i \equiv \int_{\psi_i}^{\psi_{i+1}} \frac{q}{\chi}\,\mathrm{d}\psi \Bigg/\int_{\psi_i}^{\psi_{i+1}} \frac{1}{\chi}\,\mathrm{d}\psi
    \label{macrocell_average}
\end{equation}
This prescription preserves (by construction) the integrated quantities $\int q\,\mathrm{d}m = 2\pi\int (q/\chi)\,\mathrm{d}\psi$ for each macrocell, and produces the 1D arrays $s_i$, $\chi_i$, $\ell_i$ and $\psi_i$ used by the Hungarian algorithm.

\mbox{\textit{Step 6: 1D equilibrium.}} Finally, we obtain a 1D equilibrium corresponding to the simulation state by solving Eq.~\eqref{Psigma_rsigma} with $\sigma = \mathrm{id}$ the identity permutation.

\section{Details of the numerical simulations}

Here we describe in detail the properties of the numerical simulations presented in this work. For reproducibility, we give all numerical values in code units.

\textit{Numerical method.\quad}We employ the finite-volume MHD code Athena++ \cite{Stone20} to solve the axisymmetric MHD equations with sources and sinks on a regular Cartesian grid in the poloidal plane. The domain has 288 cells in the radial direction, which extends from $r_{\mathrm{min}}=0.5$ to $r_{\mathrm{max}}=3.0$, and employs zero-stress, zero-normal-velocity boundary conditions (see below for the treatment of thermodynamic quantities at this boundary), and 352 cells in the axial direction, which extends from $z_{\mathrm{min}}=0$ to $z_{\mathrm{max}}=3.0$ and is periodic. The equations are the mass-continuity equation,
\begin{equation}
    \frac{\dd \rho}{\dd t} = -\rho\bnabla \bcdot \bu_{\mathrm{pol}} + \mathcal{S}_{\rho},\label{eq:mass-appendix}
\end{equation}
poloidal-momentum equation,
\begin{multline}
    \rho \frac{\dd \bu_{\mathrm{pol}}}{\dd t} = -\bnabla \left(p + \frac{B^2}{2}\right) + \left(- \frac{B^2}{r} + \frac{\rho u_{\phi}^2}{r}\right)\boldsymbol{\hat{r}} \\-\rho\frac{\bu_{\mathrm{pol}}}{\tau_{\rm drag,pol}} + \rho\nu \left( \nabla^2 \bu_{\mathrm{pol}} + \frac{1}{3} \bnabla \left( \bnabla \bcdot \bu_{\mathrm{pol}} \right ) \right ),\label{eq:momentum-appendix}
\end{multline}
azimuthal-momentum equation,
\begin{equation}
    \frac{\dd u_{\phi}}{\dd t} = - \frac{u_{\phi}}{\tau_{\mathrm{drag,\phi}}}w_{\mathrm{in}}(r) - \frac{u_{\phi}-u_{\phi,0}}{\tau_{\mathrm{out}}}w_{\mathrm{out}}(r)
    + \mathcal{S}_{u_\phi} + \nu \nabla^2 u_\phi, \label{eq:utheta-appendix}
\end{equation}
azimuthal induction equation,
\begin{equation}
    \frac{\dd B}{\dd t} = -B\bnabla \bcdot \bu_{\mathrm{pol}} + \eta \bnabla^2 B,\label{eq:induction-appendix}
\end{equation}
and thermal-energy equation,
\begin{multline}
    \frac{\dd }{\dd t}\frac{p}{\gamma -1 }= - \frac{\gamma}{\gamma -1 }p\bnabla \bcdot \bu_{\mathrm{pol}} + R_{\mathrm{heat}}w_{\mathrm{heat}}(r) + \mathcal{S}_{p} \\ +\frac{\rho u_{\rm pol}^2}{\tau_{\rm drag,pol}}+ \rho \nu \left( 2 e_{ij} e_{ij} - \frac{2}{3} \left( \bnabla \bcdot \bu \right )^2 \right ) \\+ \eta \left| \bnabla \times (B \hat{\boldsymbol{\phi}}) \right|^2 + \rho \kappa \nabla^2 \left( \frac{p}{\rho} \right ),\label{eq:thermal-appendix}
\end{multline}
where
\begin{equation}
    e_{ij} = \frac{1}{2} \left( \frac{\p u_i}{\p x_j} + \frac{\p u_j}{\p x_i} \right )
\end{equation}is the rate-of-strain tensor, $u_{\mathrm{pol}}\equiv |\bu_{\rm pol}|$, and the quantities $\mathcal{S}_i$ are boundary sponges, described below. We take the diffusion coefficients appearing in the above equations to be ${\nu = \eta = (\gamma - 1)\kappa = 3\times 10^{-7}}$; the factor $\gamma -1$ ensures stability against double-diffusion-like instabilities.

\label{app:ic}

\textit{Initialization in reference equilibrium.\quad}We initialize the simulations from a one-dimensional reference equilibrium profile in $r$. We choose this equilibrium to be the one with spatially constant $s=s_0$, $\chi=\chi_0$ and $\ell = \ell_0$. These constants are fixed by choosing ${p_0(\Rh) = 1.25\times10^{-3}}$, ${\rho_0(\Rh)=1.0}$, ${\mathrm{Ma}_{\phi,0}(\Rh)=1.0}$ and ${B_0(\Rh)=0.5}$, where ${\Rh = 1.0}$ is the radius about which the heating term is centered. Then,
\begin{align}
    \chi_0 &= \frac{B_0(\Rh)}{\rho_0(\Rh) \Rh},\\\quad s_0 &= \frac{p_0(\Rh)^{1/\gamma}}{\rho_0(\Rh)},\\ \ell_0&= \Rh \mathrm{Ma}_{\phi,0}(\Rh) c_{s,0}(\Rh),
\end{align}where $c_{s,0}(\Rh)=\sqrt{{\gamma p_0(\Rh)}/{\rho_0(\Rh)}}$.

For constant $s$, $\chi$, $\ell$, the equilibrium condition
\begin{equation}
    \frac{\dd}{\dd r}\left(p+\frac{B^2}{2}\right)=-\frac{B^2}{r}+\frac{\rho u_{\phi}^2}{r}
\end{equation}becomes
\begin{equation}
  \frac{{\rm d}y}{{\rm d}r}
  =
  \frac{
    \ell_0^2 y/r^5
    + 2\gamma s_0 y^\gamma r^{-2\gamma-1}
  }{
    \gamma s_0 y^{\gamma-1} r^{-2\gamma} + \chi_0 y/r^2
  },
  \label{eq:y-ode}
\end{equation}where $y(r)= rB_0(r)$. We solve Eq.~\eqref{eq:y-ode} by numerical integration (forward and backward from $r=\Rh$) subject to $y(\Rh)= \Rh B_0(\Rh)$.

\textit{Initial poloidal velocity.\quad}To break translational symmetry in the axial direction, we construct a divergence-free initial poloidal velocity $\delta \bu$ from an axisymmetric streamfunction $\Phi(r,z)$ via
\begin{equation}
    \delta u_r = \frac{1}{r}\,\partial_z \Phi,\quad \delta u_z = -\frac{1}{r}\,\partial_r \Phi,\quad \delta u_\phi = 0,
\end{equation}where
\begin{multline}
  \Phi(r,z)
  = A\,w(r)\sum_{p=1}^{p_{\max}}\sum_{n=1}^{n_{\max}}\frac{1}{k_{pn}}\times\\
    \sin\!\left[p\pi x(r) + \frac{2\pi n}{L_z}(z-z_{\min}) + \phi_{pn}\right],
  \label{eq:psi-seed}
\end{multline}
$x(r)\equiv (r-r_{\rm min})/(r_{\rm max}-r_{\rm min})\in[0,1]$,
${L_z=z_{\max}-z_{\min}}$, $p_{\mathrm{max}}=10$, $n_{\mathrm{max}}=10$, and the phases $\phi_{pn}$ are pseudorandom numbers. Because the mode set excludes $n=0$,
the perturbation has zero mean in $z$ and thus injects no net axial momentum. We taper the streamfunction by a radial window function
\begin{equation}
  w(r)=\tanh\!\left(\frac{r-r_{\rm min}}{r_{\rm buf}}\right)
       \tanh\!\left(\frac{r_{\rm max}-r}{r_{\rm buf}}\right),
\end{equation}where $r_{\rm buf} = 0.2(r_{\rm max}-r_{\rm min})$, to ensure the perturbation turns off smoothly near both radial boundaries. The factor $1/k_{pn}$, where
\begin{equation}
  k_{pn} = \sqrt{\left(\frac{p\pi}{r_{\rm max}-r_{\rm min}}\right)^2
           + \left(\frac{2\pi n}{L_z}\right)^2}.
\end{equation}is the mode wavenumber, weights the streamfunction so that the resulting velocity perturbation has approximately equal power per mode (prior to the radial taper). Finally, we choose the normalization of the streamfunction $A$ such that the root mean square of the velocity field (in volume),  $\delta {u}_{\mathrm{rms}}$, satisfies $\delta {u}_{\mathrm{rms}}^2=10^{-3}c_{s,0}(\Rh)^2$.

\begin{figure}
    \centering
    \includegraphics[width=1.0\columnwidth]{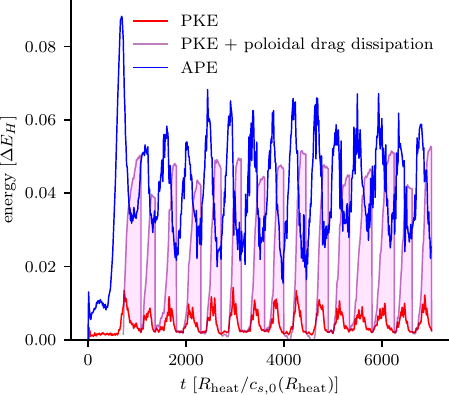}
    \caption{Same as panel (a) of Fig.~\ref{fig:cycle} in the main text, but also showing the integrated energy dissipated by the poloidal drag [i.e., the volume and time integral of the term involving $\tau_{\rm drag,pol}$ in Eq.~\eqref{eq:thermal-appendix}].}
    \label{fig:drag_fig}
\end{figure}

\textit{Poloidal drag:\quad}The $-\rho\boldsymbol{u}_{\rm pol}/\tau_{\rm drag,pol}$ term in Eq.~\eqref{eq:momentum-appendix} describes a drag on the poloidal velocity with characteristic timescale $\tau_{\rm drag,pol}=10^3$ in the simulations presented in this work. The term involving $\tau_{\rm drag,pol}$ in \eqref{eq:thermal-appendix} ensures that reduction in kinetic energy due to the drag leads to a corresponding increase in thermal energy: the drag is conservative overall. The energy dissipated by the poloidal drag is shown in Fig.~\ref{fig:drag_fig}.

\textit{Toroidal drag:\quad}The $- u_{\phi}w_{\mathrm{in}}(r)/{\tau_{\mathrm{drag,\phi}}}$ term in Eq.~\eqref{eq:utheta-appendix} describes a drag on the toroidal velocity with timescale ${\tau_{\mathrm{drag,\phi}}=1.6\times 10^{4}}$. The window function $w_{\mathrm{in}} = 1- H(r-\Rped)$, where $H(x)$ is the Heaviside (step) function and $\Rped=1.25$.

\textit{Toroidal spin-up:\quad}The $({u_{\phi}-u_{\phi,0}})w_{\mathrm{out}}(r)/{\tau_{\mathrm{out}}}$ term in Eq.~\eqref{eq:utheta-appendix} describes a relaxation of the toroidal velocity to the reference profile with timescale ${\tau_{\mathrm{out}}=1.6\times 10^{4}}$. The corresponding window function is ${w_{\mathrm{out}} = H(r-\Rped)}$.

\textit{Plasma heating:\quad} Heating is implemented as an explicit source in Eq.~\eqref{eq:thermal-appendix} localized near
$r=\Rh=1.0$ with Gaussian profile
\begin{equation}
  w_{\mathrm{heat}}(r) = \exp\!\left[-\frac{(r-\Rh)^2}{2\Delta r_{\mathrm{heat}}^2}\right],
\end{equation}where $\Delta r_{\mathrm{heat}}=0.05$.

\textit{Sponges:} We enforce sponge regions close to the inner and outer boundaries (denoted core and wall, respectively, in what follows) via
\begin{equation}
    \left(\frac{\dd Q_i}{\dd t}\right)_{\mathrm{sponges}} = -\frac{Q_{i}-Q_{i0}}{\tau^{Q_i}_{\mathrm{core}}}w_{in}(r) -\frac{Q_{i}-Q_{i0}}{\tau_{\mathrm{wall}}^i}w_{\rm out}(r)
\end{equation}for $Q_i = (s,\chi,\ell)_i$ and $Q_{0i} = (s_0,\chi_0,\ell_0)_i$. With the appropriate change of variables, the expressions for $\mathcal{S}_{\rho}$, $\mathcal{S}_{u_\phi}$ and $\mathcal{S}_p$ in Eqs.~\eqref{eq:mass-appendix}, \eqref{eq:utheta-appendix} and~\eqref{eq:thermal-appendix} are
\begin{equation}
    \mathcal{S}_{\rho}=-\rho\frac{\chi-\chi_0}{\chi}\left[\frac{w_{\mathrm{core}}(r)}{\tau^{\chi}_{\mathrm{core}}}+\frac{w_{\mathrm{wall}}(r)}{\tau^{\chi}_{\mathrm{wall}}}\right],
\end{equation}
\begin{equation}
    \mathcal{S}_{u_\phi}= -\left(u_\phi-u_{\phi,0}\right)\left[\frac{w_{\mathrm{core}}(r)}{\tau^{\ell}_{\mathrm{core}}}+\frac{w_{\mathrm{wall}}(r)}{\tau^{\ell}_{\mathrm{wall}}}\right],
\end{equation}
\begin{multline}
    \mathcal{S}_p= -\frac{\gamma p}{\gamma -1} \Bigg\{\frac{s-s_0}{s}\left[\frac{w_{\mathrm{core}}(r)}{\tau^{s}_{\mathrm{core}}}+\frac{w_{\mathrm{wall}}(r)}{\tau^{s}_{\mathrm{wall}}}\right]\\
    -\frac{\chi-\chi_0}{\chi}\left[\frac{w_{\mathrm{core}}(r)}{\tau^{\chi}_{\mathrm{core}}}+\frac{w_{\mathrm{wall}}(r)}{\tau^{\chi}_{\mathrm{wall}}}\right]\Bigg\}.
\end{multline}The inner and outer window functions are, respectively,
\begin{equation}
    w_{\rm wall}(r)=\frac{1}{2}\left[1+\tanh\!\left(\frac{r-r_{\rm wall}}{\Delta r_{\rm wall}}\right)\right]
\end{equation}and 
\begin{equation}
    w_{\rm core}(r)=1-H(r-r_{\mathrm{core}})
\end{equation}where $H(x)$ is the Heaviside (step) function. In the simulations presented in the main text, we employ $r_{\rm wall} = 2.5$,
$\Delta r_{\rm wall} = 0.1$, $r_{\mathrm{core}}=0.9$, ${\tau^{s}_{\mathrm{core}}=\tau^{\chi}_{\mathrm{core}}=500}$,  $\tau^{\ell}_{\mathrm{core}}=\infty$ (i.e., no angular-momentum sponge at the core) and ${\tau^{s}_{\mathrm{wall}}=\tau^{\chi}_{\mathrm{wall}}=\tau^{\ell}_{\mathrm{wall}}=10}$.

\textit{Axial boundary conditions:\quad}We employ periodic boundary conditions in the axial ($z$) direction.

\textit{Radial boundary conditions:\quad}In the radial direction, we employ zero-stress (i.e., $u_z$ symmetric about boundary face) and zero-normal-velocity ($u_r$ antisymmetric about boundary face) boundary conditions for the poloidal velocity, and zero-gradient boundary conditions for each of the Lagrangian invariants $\bQ=(s, \chi, \ell)$---we copy the values of these invariants from the last active cell into the ghost cells. Given a ghost-cell pressure $p_{\rm gh}(r,t)$ (defined below), the invariants uniquely determine the remaining ghost-cell primitives, $\rho_{\mathrm{gh}}$, $B_{\mathrm{gh}}$ and $u_{\phi,\mathrm{gh}}$.

\textit{Pressure:\quad}At each radial boundary, we calculate the ghost-cell pressure from an analytic equilibrium solution anchored to the boundary-adjacent active cell, multiplied by a time-dependent prefactor, see below. The leading-order low-$\beta$ force balance gives $p\propto r^{-2\gamma}$,
so we write
\begin{equation}
  p_{\rm gh}(r) = (1+A) \,p_{\rm a} \left(\frac{r}{r_{\rm a}}\right)^{-2\gamma} f(r),\label{eq:pressure_analytic_appendix}
\end{equation}
where $(r_{\rm a},p_{\rm a},\rho_{\rm a},u_{\phi{\rm a}},B_{\rm a})$ are taken from the reference active cell. The factor $f(r)=1+\lambda(r-r_{\rm a})$, where $\lambda$ is chosen to satisfy the equilibrium condition at finite $\beta$ to leading order in $r-r_{\rm a}$, which requires
\begin{equation}
  \lambda=\frac{1}{r_{\rm a}}\,\frac{\gamma\beta_{\rm a}}{\gamma\beta_{\rm a}+2}\,
  \left(2\gamma + \frac{\rho_{\rm a} u_{\phi{\rm a}}^2}{p_{\rm a}}\right),
\end{equation}where $\beta_{\rm a}={2p_{\rm a}}/{B_{\rm a}^2}$.

\textit{Pressure controller:\quad} Over the $\sim 10^5$ Alfv\'{e}nic timescales for which we run the simulation, the approximate nature of Eq.~\eqref{eq:pressure_analytic_appendix} leads to a slow, systematic mass flux through the radial
boundaries. We control this by adjusting the wall-pressure normalization~$A$ in response to measurements of the Riemann mass flux $\mathcal{F}_\rho$. At each timestep we evolve $A_{\pm}$ at the inner ($-$) and outer ($+$) radial boundaries according to
\begin{equation}
  \frac{{\rm d}A_{\pm}}{{\rm d}t}=-\frac{A_{\pm}-A_{\pm}^\star(t)}{\tau_{{A},\pm}},
\end{equation}where $\tau_{{A},\pm}=1$ is a relaxation timescale and
\begin{equation}
  A_{\pm}^\star(t)=\pm\left[G_{\pm}\,e_{\pm}(t)+\hat b_{\pm}(t)\right]\label{Astar}
\end{equation}is the normalization target, which is chosen so as to minimize the surface-averaged Riemann-solver mass flux at the inner and outer boundaries. $A_{\pm}^\star(t)$ is composed of two parts. The error term is $G_{\pm}\,e_{\pm}(t)$, where $G_{\pm}=30$ is a constant and $e_{\pm}(t)$ is the normalized Riemann flux
\begin{equation}
    e_{\pm}(t)=\frac{\left\langle \mathcal{F}_\rho \right\rangle_{r=r_{\pm}}}{\left\langle \rho_{\rm a} c_{s,{\rm a}} \right\rangle_{r=r_{\pm}}},
\end{equation}where $c_{s,{\rm a}}^2=\gamma p_{\rm a}/\rho_{\rm a}$ and $\langle\cdot\rangle$ denotes an average over boundary faces (the averages in the denominator are taken over the boundary-adjacent
active cells at each wall). The signs of the feedback in Eq.~\eqref{Astar} mean that a positive (negative) $e_{\pm}(t)$ increases (decreases) the pressure gradient at the boundary, resisting the mass flux. The bias term $\hat b_{\pm}(t)$ in Eq.~\eqref{Astar} integrates the
normalized error to remove a steady (time-averaged) offset
\begin{equation}
  \frac{{\rm d}\hat b_{\pm}}{{\rm d}t}=\frac{\,e_{\pm}(t)}{\tau_{{\rm b}}},
\end{equation}where $\tau_{{\rm b}}=20$.

\end{document}